\documentclass[sigconf,screen,natbib=false]{acmart}
\usepackage[utf8]{inputenc}
\usepackage{booktabs}
\usepackage[maxbibnames=10]{biblatex}
\usepackage{hyperref}
\usepackage[scaled=0.8]{beramono}
\usepackage{graphicx}
\usepackage{multirow}
\usepackage{bigstrut}
\usepackage{makecell}
\usepackage{pifont}
\usepackage{xspace}
\usepackage{enumitem}

\setcopyright{acmcopyright}
\copyrightyear{2018}
\acmYear{2018}
\acmDOI{XXXXXXX.XXXXXXX}
\acmConference[Conference acronym 'XX]{Make sure to enter the correct
  conference title from your rights confirmation emai}{June 03--05,
  2018}{Woodstock, NY}
\acmPrice{15.00}
\acmISBN{978-1-4503-XXXX-X/18/06}
\definecolor{darkblue}{rgb}{0,0,0.3}
\hypersetup{
	pdftitle={Security, Privacy, and Decentralization in Web3},
	pdfauthor={Philipp Winter, Anna Harbluk Lorimer, Peter Snyder, and Benjamin Livshits},
	pdfkeywords={ethereum, defi, privacy, security},
	colorlinks=true,
	urlcolor=darkblue,
	linkcolor=darkblue,
	citecolor=darkblue,
}

\newcommand{\point}[1]{\par\smallskip{\noindent\bf#1:}}

\begin{document}

\title{Security, Privacy, and Decentralization in Web3}

\author{Philipp Winter}
\affiliation{
  \institution{Brave Software}
  \country{}
  \city{}
}

\author{Anna Harbluk Lorimer}
\affiliation{
  \institution{Brave Software}
  \country{}
  \city{}
}

\author{Peter Snyder}
\affiliation{
  \institution{Brave Software}
  \country{}
  \city{}
}

\author{Benjamin Livshits}
\affiliation{
  \institution{Brave Software}
  \country{}
  \city{}
}
\affiliation{
  \institution{Imperial College London}
  \country{}
  \city{}
}

\begin{abstract}
Much of the recent excitement around decentralized finance (DeFi) comes from hopes that DeFi can be a secure, private, less centralized alternative to traditional finance systems.  However, people moving to DeFi sites in hopes of improving their security and privacy may end up with less of both as recent attacks have demonstrated.

In this work, we improve the understanding of DeFi by conducting the first Web measurements of the security, privacy, and decentralization properties of popular DeFi front ends.  We find that DeFi applications---or \emph{dapps}---suffer from the same security and privacy risks that frequent other parts of the Web but those risks are greatly exacerbated considering the money that is involved in DeFi.
Our results show that a common tracker can observe user behavior on over~56\% of websites we analyzed and many trackers on DeFi sites can trivially link a user's Ethereum address with \textsc{pii} (e.g., user name or demographic information), or phish users by initiating fake Ethereum transactions.  Lastly, we establish that despite claims to the opposite, because of companies like Amazon and Cloudflare operating significant Web infrastructure, DeFi as a whole is considerably less decentralized than previously believed.
\end{abstract}

% ACM-specific stuff.


\received{20 February 2007}
\received[revised]{12 March 2009}
\received[accepted]{5 June 2009}

\maketitle

\section{Introduction}
\label{sec:intro}

We are witnessing a movement in finance whose goal is to replace long-standing institutions with code. More than~25 billion U.S. dollars are currently locked in decentralized finance~(DeFi)~\cite{defipulse}. Guided by the principles of decentralization and self-custody, DeFi applications implement numerous instruments from traditional finance like insurance, loans, or exchanges which are now steered by algorithms rather than human intervention.  DeFi applications differ significantly in their architecture from traditional websites: Instead of a complex Web front end that communicates with a database back end, DeFi applications expose a lightweight front end that interacts with a blockchain-based back end using wallet software that runs in the user's browser. This unorthodox design---coupled with the substantial and ever-increasing amount of money pouring into DeFi---raises several questions related to security and privacy: What is the attack surface that attackers could exploit to steal funds?  How well do DeFi sites protect the user's financial information?  What is the role of third-party trackers?  This work sets out to answer these questions. While smart contract security has been studied extensively~\cite[\S~5.1]{Werner21a}, the above questions have received little attention. 

We begin by compiling a list of~78 popular DeFi sites and proceed to study the problem through the lens of well-established Web security and privacy methods. In particular, we identify what third-party providers DeFi sites use, what user data those providers can obtain, and how those providers could misbehave. 

Our results show that 66\% of DeFi sites rely on scripts from third-party providers and \emph{more than half} of all DeFi sites rely on analytics scripts from Google, which gives the advertising company substantial insight into how users interact with DeFi sites.  We also find that 17\% of DeFi sites leak the user's Ethereum address to third parties, which constitutes a significant privacy issue.
We also look at problems that are caused by the architectural design of DeFi applications:  The scripts that a DeFi site embeds are able to interact with the user's wallet \textsc{api}, which facilitates phishing attacks.
%
% Front-end decentralization.
Last, we find that DeFi sites frequently fall short of being truly decentralized, jeopardizing security, governance, and reliability.  For example, Cloudflare hosts almost half of all sites and therefore has comprehensive data on how users interact with DeFi applications.  Motivated by the infamous event-stream vulnerability~\cite{EventStream}, we study the software dependencies of DeFi front ends in an attempt to understand what software packages are popular, and to get a sense of the potential for injecting malicious code in any of these dependencies.

\point{Contributions}
This paper makes the following contributions:
\begin{enumerate}
    \item We compiled a list of~78 popular DeFi sites and built a JavaScript crawler that visits these sites and extracts the Web requests that they make before and after a user connects her wallet to the DeFi site.  We make our code and this data set publicly available.

    \item We analyzed the aforementioned data set for privacy and security problems.  Our analysis revealed cross-site Ethereum address leaks and it showed that Google can track Web3 users across the marketing funnel and do conversion analysis.  We also studied the dependencies of DeFi front ends (finding that the average front end relies on a median of 53 dependencies) and the underlying hosting infrastructure of DeFi front ends. 

    \item We performed measurements of the different dimensions of decentralization within our data set and concluded that the lack of decentralization is a surprising Achilles heel of a domain that claims to be significantly decentralized.  For instance, we find that Cloudflare and \textsc{aws} are responsible for more than~80\% of DeFi site front ends.
\end{enumerate}

\point{Paper organization}
In the rest of this work, Section~\ref{sec:background} provides background on DeFi, followed in Section~\ref{sec:issues} by an explanation of the privacy and security issues we discovered.  We then measure the prevalence of these issues on popular DeFi sites in Section~\ref{sec:measurements}.  Section~\ref{sec:discussion} discusses this work's limitations and makes recommendations to both users and DeFi developers.  We conclude this work in Section~\ref{sec:related} by contrasting it with past research and summarize our findings in Section~\ref{sec:conclusions}.
\section{Background}
\label{sec:background}

\subsection{Web Tracking}

Users are typically tracked on the Web by third parties that use either explicit or implicit information to track users across sites. Historically, third-party cookies have been a popular way of tracking users but browsers like Firefox and Safari have recently started blocking third-party cookies by default~\cite{firefox-cookies,safari-cookies}, which makes them a less effective tracking vector. Numerous alternative tracking vectors exist though, like browser fingerprinting, whose idea is to uniquely identify users based on the nuances of their browser's configuration, e.g., screen resolution, available fonts, or installed extensions.

\subsection{Ethereum}

Ethereum's native currency is called Ether and users often manage (some of) their Ether in software wallets that are implemented as browser extensions.  These extensions store the user's private key\footnote{Note that one can also store the private key in a hardware wallet and manage it via a software wallet.} and provide a UI for managing the user's funds.

Unlike Bitcoin, Ethereum implements an account-based model, similar to bank accounts. All transactions from and to a given Ethereum account are eternalized on the blockchain and can easily be linked to the user’s account (but not necessarily their real-world identity), which means that at best, users enjoy pseudonymity.  It is however straightforward to create multiple Ethereum accounts to compartmentalize one’s financial activity.  While privacy layers have been proposed on top of Ethereum, their usage remains scarce~\cite{Aztec,StarkEx,Tornado}. 

Users are encouraged to keep their Ethereum address secret because knowing an address reveals the owner’s funds (which may paint a target on one’s back) and transaction history (which may reveal sensitive relationships). Regardless, many users freely share their Ethereum address online and, of course, users needs to reveal their address to some parties (e.g., Ethereum nodes) to use the network meaningfully. Given Ethereum’s emerging financial ecosystem, online advertisers have an incentive to obtain users' Ethereum address for targeted advertisement; this advertisement can include traditional e-commerce if wallet transactions are used for online marketplaces or, more likely, crypto-focused ads for \textsc{nft}s or services such as crypto-friendly debit cards, for instance. 

\subsection{Decentralized Finance Software Stack}

\begin{figure}[t]
    \centering
    \includegraphics[width=\linewidth]{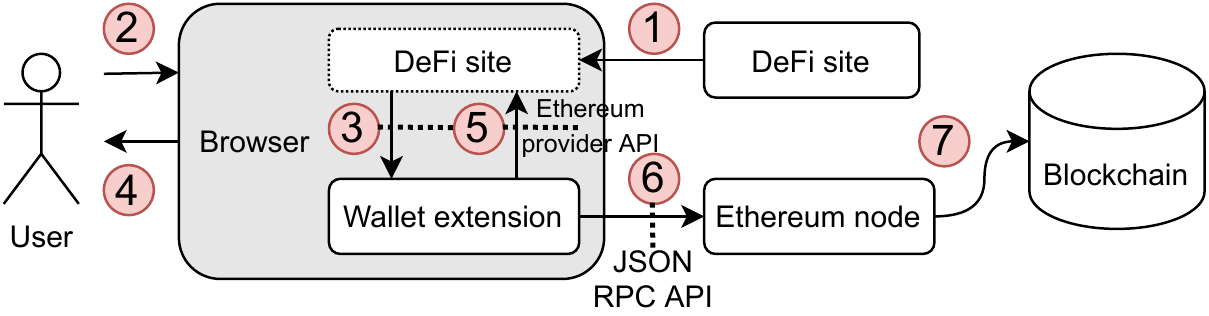}
    \caption{The conceptual flow of DeFi sites: \ding{182} Users visit a DeFi site, \ding{183} click the ``connect wallet'' button, \ding{184} which prompts the DeFi site to ask the user's wallet for permission. \ding{185} The user then grants permission in the UI, \ding{186} after which the DeFi site can access the user's Ethereum account, and create transactions that make it \ding{187} via an Ethereum node \ding{188} to the blockchain.}
    \label{fig:defi-architecture}
\end{figure}

Centralized exchanges, like Coinbase, run counter to the philosophy of decentralization that is at the core of cryptocurrencies, which is why decentralized alternatives have emerged.  Typically referred to as DeFi, these applications allow users to invest in liquidity pools, exchange tokens, send payments, or lend money. A particularly popular example is Uniswap, a site that allows users to swap \textsc{erc}-20 tokens and invest in liquidity pools---tasks that have historically only been offered by centralized exchanges like Coinbase. 

As illustrated in Figure~\ref{fig:defi-architecture}, DeFi applications essentially consist of a Web interface that bridges the gap between a user’s Ethereum wallet (e.g., MetaMask~\cite{metamask}) and the DeFi application’s smart contract.  DeFi sites therefore need to interact with the Ethereum blockchain \emph{and} the user's wallet. Both types of interactions can (but don't need to) take place via the Ethereum provider \textsc{api}~\cite{EthProvAPI}---a JavaScript \textsc{api} that the MetaMask extension injects into the DeFi site's \textsc{dom}. The DeFi site can then interact with the \textsc{api} via the \texttt{window.ethereum} JavaScript object. A subset of the Ethereum provider \textsc{api} is handled directly by MetaMask, e.g. the signing of transactions. The remaining \textsc{api} calls are \emph{not} handled by MetaMask and forwarded to an Ethereum node, e.g. to an Ethereum-as-a-service provider like Infura~\cite{infura}.  Note that a DeFi site does not have to rely on MetaMask to interact with an Ethereum node; it could simply talk to an Ethereum node directly---many DeFi sites do---and limit the interaction with the Ethereum provider \textsc{api} to the signing of transactions. Unlike MetaMask, Ethereum nodes expose a \textsc{json}-based \textsc{rpc} interface. This interface consists of several dozen functions to call smart contracts, fetch gas prices, or obtain the number of the most recent Ethereum block.

Note that MetaMask makes available some of its \textsc{api} functions only \emph{after} the user gave permission---typically by manually clicking the ``connect wallet'' button\footnote{E.g., via the following JavaScript call:\\\texttt{window.ethereum.request(\{method: 'eth\_requestAccounts'\})}} (to prevent unauthorized sites from accessing the user's Ethereum account information~\cite{EIP-1102}), which prompts the DeFi site to ask MetaMask for permission, followed by a UI dialog asking the user to confirm the DeFi site's request. Once the user gave permission, the DeFi site is able to access the user's Ethereum address and balance, and create transactions (that still need to be signed off by the user). 
For a more comprehensive overview of DeFi, refer to Werner et al.'s 2021 arXiv report~\cite{Werner21a}.
\section{Issues and Attacks in Web3 Sites}
\label{sec:issues}

In this section, we briefly outline the privacy and security issues we discovered among DeFi sites by using the example of 1inch.exchange (in short: 1inch).  None of those issues are new---we could summarize them as ``the past comes back to bite us again.''  What makes those issues relevant is the nature of DeFi sites, which differs from traditional websites in that (i) substantial amounts of money are involved and all that stands between a malicious DeFi site and the user's funds is often just a layer of JavaScript; (ii) a user's Ethereum address is effectively a unique, long-term identifier that is linked to the user's publicly visible financial history---ideal conditions for online tracking. A summary of issues and attacks covered below is shown in Table~\ref{tab:issue-summary}.

\point{Privacy}
1inch's landing page embeds scripts from several third parties, one of which is Google Analytics.  This reveals arguably sensitive financial browsing activity to Google because 1inch encodes in its \textsc{url}s what tokens the user is interested in exchanging, e.g. the \textsc{url} \url{https://app.1inch.io/\#/1/swap/COMP/USDC} (which makes its way to Google Analytics) reveals that the user selected the \textsc{comp} and \textsc{usdc} token to swap.  Furthermore, the use of Google Analytics also leaks the user's \emph{Ethereum address} because 1inch happens to put the user's Ethereum address in an ``event label'' on Google Analytics. This allows Google to link the user's Ethereum address with the \textsc{pii} the company likely already has about the user. Worse, Google---and other analytics providers---also play a role in other DeFi sites, making it possible to track users \emph{across sites}.

Analytics providers enjoy widespread use thanks to the convenience and insight that they provide but the nature of data they leak to third parties is highly sensitive in the context of DeFi.

\point{Security}
1inch embeds a chat widget that allows users to contact 1inch support staff. The widget is provided by a third party and consists of JavaScript that is embedded in 1inch’s first party context and therefore has \emph{full control} over 1inch’s \textsc{dom}. That by itself is not surprising---virtually all major websites embed scripts but once the user connects her MetaMask wallet to 1inch, both 1inch \emph{and} the embedded chat widget are able to interact with the Ethereum provider \textsc{api} that’s injected by MetaMask. Crucially, the chat widget (or whoever compromised its distribution infrastructure) can modify 1inch's \textsc{dom} to phish the user in an attempt to steal funds, or directly make a transaction via the user's wallet. While this transactions would have to be approved by the user, we argue that a well-crafted transaction at the right time would fool many users.  First-party script inclusion always brings with it a certain risk but we believe that this risk is \emph{highly elevated} in the context of DeFi considering the amount of money that is at stake.

% TODO: Maybe mention Node.js dependency analysis?  1inch doesn't publish its front end code though.

\point{Excessive centralization}
1inch's front end is made available via Cloudflare's \textsc{cdn} and so are most other DeFi sites. The issue here is twofold: (i) centralizing front ends among only a few hosting providers can lead to a substantial amount of funds being effectively unavailable should an outage occur; and (ii) centralized hosting providers gain (and potentially monetize) a lot of insight into how users interact with DeFi.  Besides, governments can coerce centralized hosting providers into ceasing operations which is not difficult to imagine considering how the U.S. government sanctioned the popular mixer Tornado Cash~\cite{tornado-sanctions}.

\begin{table}[t]
\centering\setlength{\tabcolsep}{5pt}
\caption{A summary of the potential attacks and issues this paper presents.}
\label{tab:issue-summary}
\begin{tabular}{lp{4.5cm}}
\toprule
\bf Issue type & \bf Description \\
\midrule
Privacy (\S~\ref{sec:addr-leaks})&
\begin{minipage}[t]{\linewidth}
    \begin{itemize}[leftmargin=*]
        \item Ethereum address leaks to third parties.
        \item Third-party trackers can track DeFi users across sites.
    \end{itemize}
\end{minipage} \\

& \\

Security (\S~\ref{sec:dependencies}, \S~\ref{sec:supply-chain}) &
\begin{minipage}[t]{\linewidth}
    \begin{itemize}[leftmargin=*]
        \item DeFi sites heavily rely on third-party scripts (and sometimes iframes) that could phish the user.
        \item Front ends rely on a large number of Node dependencies.
    \end{itemize}
\end{minipage} \\

& \\

Decentralization (\S~\ref{sec:centralization}) &
\begin{minipage}[t]{\linewidth}
    \begin{itemize}[leftmargin=*]
        \item Front ends and hosting infrastructure are mostly centralized.
    \end{itemize}
\end{minipage} \\

\bottomrule
\end{tabular}
\end{table}
\section{Measurements}
\label{sec:measurements}

Having introduced potential privacy and security issues, we now turn to understanding how common those issues are.  How many DeFi sites exhibit one or more of those issues?  How common is the presence of Google Analytics?

We begin by discussing our measurement method (\S~\ref{sec:method}), followed by an analysis of how often a user's Ethereum address leaks to third parties (\S~\ref{sec:addr-leaks}).  We then take a step back and determine what third parties DeFi sites rely on and what this implies (\S~\ref{sec:dependencies}).  Next, we study what software packages DeFi sites depend on (\S~\ref{sec:supply-chain}) and the degree of decentralization among DeFi hosting providers (\S~\ref{sec:centralization}).

\subsection{Data Set and Method}
\label{sec:method}

We begin by compiling a list of DeFi sites to inspect from DeFi Pulse~\cite{defipulse}, which lists the top DeFi sites ranked by ``total value locked,'' i.e. the amount of money that is currently ``locked'' inside the respective smart contracts---an apt proxy for the popularity of DeFi sites.  We added the top 50 sites as of 2021-08-26 and augmented the list with 26 sites that we found of interest, resulting in a list of 78 DeFi sites, shown in full in Table~\ref{tab:defi-urls} in Appendix~\ref{sec:defi-urls}.  We then manually turned the list of domains into \textsc{url}s so that when clicked, the browser lands directly on the page that asks users to connect their wallet, e.g. we turned compound.finance into https://app.compound.finance.

In the next step, we set out to visit each site and record the requests that it makes.  To facilitate that, we built a Puppeteer-based crawler that spawns an instance of Google Chrome~92.0.4515.159 on Linux, sequentially visits all \textsc{url}s in our DeFi list for 30 seconds, and records each request that the respective site makes.  We used a fresh Chrome profile for our experiment, installed the MetaMask 10.0.3 extension, and completed MetaMask's onboarding process, resulting in a new Ethereum address.  We are particularly interested in what requests a site makes \emph{after} the user connects her wallet, which is why we manually click on the ``connect wallet'' button once our crawler visits a new site.  It took less than one hour to complete these semi-automatic measurements.

For each site, our crawler created a \textsc{json} file that contains metadata about the respective site and a list of Web requests, each consisting of (i) the request context (e.g., the site itself, or an iframe), (ii) the requested \textsc{url}, and (iii) the type of request (e.g., a fetch, image, or script request).  Below is an example from Uniswap: A Web request to app.tryroll.com for a list of tokens.

\begin{verbatim}
  {
    "requestContext": [
      "https://app.uniswap.org/#/"
    ],
    "url": "https://app.tryroll.com/tokens.json",
    "type": "fetch"
  },
\end{verbatim}

\subsection{Ethereum Address Leaks}
\label{sec:addr-leaks}

Having recorded all requests of the most popular DeFi sites, we ask:  Do any of these requests leak our Ethereum address to third parties?  For example, did foo.finance send an \textsc{xhr} request containing our Ethereum address to bar.finance?  To answer this question, we filter our data for requests (i) whose destination has a different e\textsc{tld}+1 than the origin\footnote{We added a special case for the sites 1inch.exchange and balancer.fi because they also operate 1inch.io and balancer.finance, respectively---different e\textsc{tld}+1 domains that are run by the same organization.} and (ii) whose \textsc{url} contains our Ethereum address.

\begin{table}
\centering
\caption{DeFi sites that leaked our Ethereum address to third parties.}
\label{tab:address-leaks}
\begin{tabular}{lr}
\toprule
\bf DeFi site & \bf \# of leaks \\
\midrule
\url{yearn.finance} & 4 \\
\url{defisaver.com} & 3 \\
\url{bifi.finance} & 3 \\
\url{zerion.io} & 3 \\
\url{loopring.io} & 2 \\
\url{bancor.network} & 2 \\
\url{dodoex.io} & 2 \\
\url{sablier.finance} & 1 \\
\url{reflexer.finance} & 1 \\
\url{impermax.finance} & 1 \\
\url{1inch.io} & 1 \\
\url{jelly.market} & 1 \\
\url{rarible.com} & 1 \\
\bottomrule
\end{tabular}
\end{table}

Table~\ref{tab:address-leaks} contains the DeFi sites that leaked our Ethereum address, along with the number of leaks we found.  Our script detected that 13 out of our~78 sites~(17\%) leaked our Ethereum address to third-party domains.  Yearn issued four fetch requests to api.zapper.fi---to retrieve ``Yearn Vaults'' and token balances; DeFi Saver issued fetch requests to defiexplore.com and api.compound.finance; BiFi requested four images from heapanalytics.com whose \textsc{url} contained our Ethereum address; and Zerion issued three fetch requests to ipfs.3box.io and maker.ifttt.com.

For the most part, this behavior represents standard Web development: many sites use third-party \textsc{api}s to submit and retrieve user information.  We must however hold DeFi front ends to a higher standard, similar to traditional financial institutions, because a user's Ethereum address constitutes sensitive financial information.

% \phw{Point out that some DeFi sites are (mostly) self-contained (copound and tornado cash);  app.deversifi.com sends your Ethereum address in an API request to go.crisp.chat.}

\subsection{Cross-origin Dependencies}
\label{sec:dependencies}

Recall from Section~\ref{sec:issues} that DeFi sites---like traditional banking sites---must be particularly careful about what external scripts they embed because those scripts can see wallet balances and potentially phish the user by modifying the page's \textsc{dom}, which raises the question: how many DeFi sites embed scripts from third parties?  We answer this question by extracting script requests from our DeFi list whose destination e\textsc{tld}+1 differs from the site's origin.

Our data shows that~48 DeFi sites~(66\%) embed at least one script from a total of~34 third parties.  Table~\ref{tab:top-dependencies} shows the top ten third parties that are embedded the most.  Google's Tag Manager can be found on~28 DeFi sites, followed by Google Analytics on~21 sites.  Intercom provides a chat support widget that can be found on~8 DeFi sites.  Google's presence among DeFi sites is pervasive: Our data shows that 41 DeFi sites~(56\%) embed at least one script provided by Google.

\point{Conversion analysis}
Is Google able to monetize Alice's behavioral data as she navigates DeFi sites?  For example, several DeFi sites leak their users' Ethereum address directly to Google Analytics, giving the company the opportunity to link a user's real-world identity (which is \textsc{pii}) to her Ethereum address.  What else can Google and other third party trackers learn about Alice?

\begin{table}
\centering
\caption{Top ten third party sites whose scripts were embedded the most.}
\label{tab:top-dependencies}
\begin{tabular}{lr}
\toprule
\bf Third party & \bf \# of sites \\
\midrule
\url{googletagmanager.com} & 28 \\
\url{google-analytics.com} & 21 \\
\url{intercomcdn.com} & 8 \\
\url{intercom.io} & 8 \\
\url{airswap.io} & 6 \\
\url{cloudflareinsights.com} & 5 \\
\url{facebook.net} & 3 \\
\url{crisp.chat} & 3 \\
\url{google.com} & 2 \\
\url{gstatic.com} & 2 \\
\bottomrule
\end{tabular}
\end{table}

Advertising companies have an incentive to track users through conversion funnels---an e-commerce term that refers to a user's journey from first hearing about a product, to considering a purchase, to finally purchasing the product. Applied to the space of cryptocurrency, this could mean tracking Alice in each step as she (i) sees an ``ad'' for a currency or a token on a news site, (ii) inspects the asset's price chart on a price discovery site, and (iii) purchases the asset on a DeFi site. A concrete funnel could look as follows: Alice browses the popular news site CoinDesk where she learns about a new token that recently appreciated in value.  To learn more about the token's price performance, she visits the price listing site CoinGecko.  Having gained faith in the token's performance, Alice decides to invest in it and visits Uniswap to make the purchase. An advertisement company tracking Alice through the funnel can enrich its profile about Alice, allowing for more effective targeted advertisement.

To answer these questions, we need websites representing the top and middle part of the conversion funnel---a list of cryptocurrency news sites and of price discovery sites.  We already have a list of DeFi sites, which constitutes the bottom part of the funnel.  We obtained a list of five news sites and six price discovery sites (see Table~\ref{tab:news-price-sites}) by selecting the top Google search results for the keywords ``cryptocurrency price'' and ``crypto news.''  Equipped with three lists of sites, a subset of which Alice would traverse until she eventually conducts a transaction, we now turn to understanding what entities can track Alice.

\begin{table*}[ht]
\centering
\caption{The list of cryptocurrency news sites and price discovery sites that we use in our analysis.}
\label{tab:news-price-sites}
\begin{tabular}{l@{\hskip 20pt}l}
\toprule
\bf News sites & \bf Price discovery sites \\
\midrule
\url{https://www.coindesk.com} & \url{https://www.coingecko.com/en} \\
\url{https://cointelegraph.com} & \url{https://coinmarketcap.com} \\
\url{https://decrypt.co} & \url{https://coinranking.com} \\
\url{https://cryptonews.com} & \url{https://coincodex.com} \\
\url{https://www.theblockcrypto.com} & \url{https://coincheckup.com} \\
& \url{https://www.cointracker.io/price} \\
\bottomrule
\end{tabular}
\end{table*}

For each of the sites in our three categories, we determine the third parties (identified by their e\textsc{tld}+1) from which the sites embed scripts.  We then determine what third parties can track users in all three categories---news site, price discovery, and purchase.  Finally, we select the subset of entities that is able to track on at least~1\% of sites in each of the three categories.  In other words, we select entities that can track on at least~1\% of news sites \emph{and} on at least~1\% of price discovery sites, \emph{and} on at least~1\% of DeFi sites.  Table~\ref{tab:conversion} lists the five companies that survived our selection criteria, and the respective percentage of sites per category that they can track on.

\begin{table}[ht]
\centering
\caption{Companies whose scripts are embedded in at least 1\% of sites of each of our three funnels parts---consisting of news sites, price discovery sites, and DeFi sites.}
\label{tab:conversion}
\begin{tabular}{l@{\hskip 10pt}r@{\hskip 10pt}r@{\hskip 10pt}r}
\toprule
\bf Third party & \bf \% of & \bf \% of & \% \bf of \\
& \bf news sites & \bf price sites & \bf DeFi sites \\
\midrule
Google & 100 & 100 & 56 \\
Cloudflare & 20 & 50 & 7 \\
Facebook & 60 & 33 & 4 \\
Hotjar & 40 & 33 & 3 \\
LinkedIn & 20 & 17 & 1 \\
\bottomrule
\end{tabular}
\end{table}

What stands out is that Google is virtually omnipresent, observing \emph{all} news and price discovery sites and \emph{most} DeFi sites.  Cloudflare, Facebook, Hotjar, and LinkedIn are all only present on less than 10\% of DeFi sites, resulting in less opportunity to track users through the conversion funnel.

Our set of news, price discovery, and DeFi sites is incomplete but we don't expect a larger set to change our results significantly because of the prevalence of Facebook, Cloudflare, and especially Google. We may however see other sites take the place of the less popular Hotjar and LinkedIn.

\subsection{Complexity of Software Dependencies}
\label{sec:supply-chain}

Node's ecosystem has fallen prey to several high-impact supply chain attacks over the last few years.  The year 2018 brought the event-stream incident in which an attacker used social engineering to gain write access to the event-stream module, which the attacker later used to add a dependency to a malicious module~\cite{EventStream}. More recently, in~2021, an attacker managed to compromise the npm account of a package maintainer, and used it to publish malicious updates~\cite{Bannister2021a}.  Zahan et al. systematically studied the problem of supply chain attacks in the Node ecosystem~\cite{Zahan2021a}.

Upon inspecting the source code of DeFi front ends, we noticed that they rely heavily on Node's ecosystem.  A security issue in a front end dependency could jeopardize the funds of DeFi users if the dependency affects the way transactions are crafted.  To understand the scope of the problem, we first study the number of dependencies that DeFi front ends rely on.  We searched for the front end code of each DeFi site in our list and cloned the respective repositories of 22~(28\%) out of our~78 DeFi sites.  All front ends are built using Node.  In the next step, we determined the number of \emph{immediate} dependencies and \emph{transitive} dependencies, which includes the (recursive) dependencies of immediate dependencies.  We determined immediate dependencies by counting the number of package names in the ``dependencies'' object in the package.json file and we determined transitive dependencies by counting the number of dependencies in the yarn.lock or package-lock.json file---whichever a project used.

Our code normalizes all transitive dependencies by removing version numbers to make sure that no dependency is counted more than once.  Table~\ref{tab:dependencies} illustrates the results.  With the exception of tinlake.centrifuge.io, all DeFi sites have at least 37 direct and more than~1,000 transitive dependencies.  A large number of dependencies does increase a DeFi site's attack surface but a vulnerability in a dependency does not automatically jeopardize the security of Ethereum transactions.  For example, we inspected vulnerabilities specific to app.aave.com by running npm's audit feature over the front end code.  The audit feature found a critical vulnerability in immer~\cite{CVE-2020-28477}, one of Aave's transitive dependencies.\footnote{The dependency path consists of the npm packages\\react-scripts $\rightarrow$ react-dev-utils $\rightarrow$ immer.}

% Direct dependencies are found under the "dependencies" key in package.json.
%   cat package.json | jq '.dependencies | length'
% Transitive dependencies are either found in package-lock.json or yarn.lock:
%   get-deps.py package-lock.json
%   grep -c '^[^ #]' yarn.lock
\begin{table}[t]
\centering
\caption{The number of immediate and transitive Node dependencies of the DeFi sites whose front end was publicly available.  The last column represents the git commit ID we inspected.}
\label{tab:dependencies}
\begin{tabular}{lrrr}
\toprule
\bf DeFi site & \bf Imm. & \bf Trans. & \bf Commit ID \\
& \bf deps. & \bf deps. & \\
\midrule
% airswap.io & 46 & 1,753 & \tt 07e51bf \\
airswap.io & 50 & 2,243 & \tt 8a60e193 \\
% app.aave.com & 66 & 2,176 & \tt 409477d \\
app.aave.com & 66 & 2,176 & \tt f34f1cfc \\
% app.balancer.fi & 94 & 1,618 & \tt 83b9a8c \\
app.balancer.fi & 107 & 1,664 & \tt 0b470019 \\
% app.bancor.network & 65 & & \tt deab8f9 \\
app.bancor.network & 47 & 1,838 & \tt be27a821 \\
% app.barnbridge.com & 46 & 2,146 & \tt 0dc9b22 \\
app.barnbridge.com & 46 & 2,067 & \tt 0811cae7 \\
app.compound.finance & 53 & 1,525 & \tt 36549ad6 \\
% app.pickle.finance & 61 & & \tt c0f9206 \\
app.pickle.finance & 79 & 1,197 & \tt ccf5512f \\
% app.rari.capital & 65 & 1,962 & \tt 8da8e84 \\
app.rari.capital & 66 & 1,963 & \tt f95a64b9 \\
% app.ribbon.finance & & & \tt abbe01a \\
% Direct dependencies are in the webapp/ directory.
app.ribbon.finance & 41 & 2,115 & \tt 53ac542a \\
% app.sushi.com & & & \tt 0d7470d \\
% No direct dependencies in package.json?
app.sushi.com & n/a & 1,806 & \tt cd611221 \\
% app.tornado.cash & & & \tt e4bc945 \\
app.tornado.cash & 39 & 2,187 & \tt a83fae07 \\
% app.uniswap.org & 56 & & \tt 034b3e3 \\
app.uniswap.org & 115 & 2,389 & \tt 4806c690 \\
% dmm.exchange & 26 & & \tt dec9e5e \\
dmm.exchange & 37 & 2,597 & \tt 8a3174f6 \\
% exchange.loopring.io & 45 & & \tt a017ce8 \\
exchange.loopring.io & 54 & 3,511 & \tt 6bd6d6ab \\
impermax.finance & 44 & 2,289 & \tt bdb74ef4 \\ % Hasn't changed.
% inverse-web.vercel.app & 30 & 1,028 & \tt 3f0325c \\
inverse-web.vercel.app & 39 & 1,143 & \tt 7a77d459 \\
% oasis.app & 96 & & \tt dae853d \\
oasis.app & 113 & 3,611 & \tt 6c78d7df \\
% pancakeswap.finance & 62 & & \tt 63c94a9 \\
pancakeswap.finance & 55 & 2,602 & \tt 571b2092 \\
% saddle.exchange & 44 & 1,883 & \tt 70877c3 \\
saddle.exchange & 52 & 1,793 & \tt dee6e290 \\
% staking.synthetix.io & 49 & 1,929 & \tt 2c78607 \\
% UI code is in v2/ui/ directory.
staking.synthetix.io & 90 & n/a & \tt 7b43a7de \\
tinlake.centrifuge.io & 4 & 4,438 & \tt da02df19 \\
% yearn.finance & 58 & & \tt 60979a1 \\
yearn.finance & 63 & 2,578 & \tt c57fcb3e \\
\bottomrule
\end{tabular}
\end{table}

Next, we seek to identify dependencies that are shared across DeFi sites.  Such dependencies constitute particularly fruitful targets considering their prevalence across DeFi sites.  Table~\ref{tab:dep-intersection} shows the npm packages that are shared by at least eight of the seventeen DeFi sites whose front end code we analyzed.  Unsurprisingly, almost all DeFi sites depend on the popular user interface library React. Almost as popular is ethers~\cite{ethers}---an Ethereum wallet implementation that allows DeFi sites to connect to Ethereum nodes via wallets like MetaMask.

% TODO: Figure out what damage an ethers vulnerability could do.

\begin{table}[t]
\centering
\caption{The immediate npm dependencies (left column) and the number of DeFi sites (right column) that share the respective dependency.}
\label{tab:dep-intersection}
\begin{tabular}{p{5.5cm}r}
\toprule
\bf Immediate dependencies & \bf Shared by (\#) \\
\midrule
react & 14 \\
react-dom & 14 \\
@web3-react/core & 10 \\
ethers & 9 \\
bignumber.js & 9 \\
@web3-react/injected-connector & 9 \\
react-redux & 8 \\
lodash & 8 \\
@web3-react/walletconnect-connector & 7 \\
graphql & 7 \\
react-router-dom & 7 \\
react-scripts & 7 \\
styled-components & 7 \\
@reduxjs/toolkit & 7 \\
\bottomrule
\end{tabular}
\end{table}

\subsection{Front End Decentralization}
\label{sec:centralization}

DeFi applications are built on top of a decentralized back end---the blockchain---but the corresponding front ends are often centralized and under the full control of the organization working on the DeFi application.  At first glance, this does not look like a problem because front ends are interchangeable: Anyone can create a front end for an existing smart contract, so if the ``primary'' front end for a DeFi application were to go down, alternatives would spin up.  This however does not reflect how less technical users interact with websites.  Users are conditioned to remember canonical domains and don't do well in evaluating the trustworthiness of alternatives~\cite[\S~5.2.5]{Winter2018a}.  Besides, there is a financial incentive for third-party front ends to embed malicious code that could hijack transactions.  Non-technical users are in a poor position to avoid such scams.  There are several compelling reasons to decentralize front ends:

\begin{description}
    \item[Security] A single controlling organization could inject malicious code at will.
    \item[Governance] A single controlling organization could act against the community's will~\cite{Uniswap2021a}.
    \item[Reliability] An outage could render funds unavailable to many users not knowledgeable enough to use alternative front ends.
\end{description}

We now quantify the centralization of DeFi front ends by determining the underlying hosting providers, which gives us an idea of the above reliability attribute.  In particular, we determined the hosting provider for each of the sites on our list.  We resolved each domain in our list of DeFi sites to its corresponding IP address and queried Team Cymru's IP-to-ASN mapping tool to map an IP address to its autonomous system number~\cite{cymru-asn}.  We then extract the autonomous system name that is also provided by Team Cymru's service.  Table~\ref{tab:hosting} summarizes the results.  Cloudflare and \textsc{aws} (Amazon) hosted a significant number of the sites on our list, indicating a trend towards  front ends being centralized behind large, corporate hosting providers.

\begin{table}[t]
\centering
\caption{The providers (as well as the number and percentage) that host the 78 DeFi sites in our list.}
\label{tab:hosting}
\begin{tabular}{lrr}
\toprule
\bf Hosting provider & \bf \# of sites & \bf \% of sites \\
\midrule
Cloudflare & 34 & 44 \\
\textsc{aws} & 30 & 38 \\
Digital Ocean & 5 & 6 \\
Fastly & 3 & 4 \\
Other & 6 & 8 \\
\bottomrule
\end{tabular}
\end{table}

Of the~78 sites surveyed, more than half were hosted by Cloudflare; \textsc{aws} hosted approximately a third of the sites; Digital Ocean hosted only four, and other hosting providers covered a negligible number of the sites.  As seen with Cloudflare’s 2019 outage that resulted in an 80\% drop of Cloudflare's traffic volume~\cite{Graham-Cumming2019a}, having the majority of DeFi sites behind one provider risks a majority of services being unavailable in the event of a large outage.

Cloudflare and \textsc{aws} clearly play an important role but what amount of funds would become unavailable if one of the above hosting providers had a global outage?  We refer to DeFi Pulse's estimates to approximate an answer, listed in Table~\ref{tab:outage-cost}.  Cloudflare and \textsc{aws} together host front ends that allow the management of approximately 90 billion \textsc{usd} worth of funds.  Those funds would not be lost entirely---manual transactions or alternative Web interfaces hosted via different means would still facilitate fund management.  While uncommon, large-scale hosting provider outages are not unheard of: Cloudflare~\cite{Graham-Cumming2019a}, \textsc{aws}~\cite{Peters2020a}, and DigitalOcean~\cite{DAStatus} have all suffered substantial outages in the past.

\begin{table}[t]
\centering
\caption{The amount of funds that would become potentially unavailable if one of the hosting providers had an outage. These numbers are estimates as the exact amount of locked in value behind DeFi sites is difficult to determine.}
\label{tab:outage-cost}
\begin{tabular}{lr}
\toprule
\bf Hosting provider & \bf Affected (\textsc{usd}) \\
\midrule
Cloudflare & 61.6 B \\
\textsc{aws} & 28.1 B \\
Fastly & 2.4 B \\
Digital Ocean & 0.9 B \\
Other & 0.8 B \\
\bottomrule
\end{tabular}
\end{table}

The damage that could be done by violation of the above security and governance attributes is harder to quantify.
\section{Discussion}
\label{sec:discussion}

In this section we discuss our work's limitations (\S~\ref{sec:limitations}) and issue recommendations for both DeFi developers and users (\S~\ref{sec:recommendations}).

\subsection{Limitations}
\label{sec:limitations}

% We can't detect all address leaks.
Our measurement method from Section~\ref{sec:method} revealed some Ethereum address leaks to third parties but it is unable to reveal \emph{deliberately disguised} address leaks.  For example, unsophisticated address encoding schemes like Base64 could have evaded our detection method.

% We don't have data on the long tail of DeFi sites.
In this paper, we focused on 78 DeFi sites---most of which are among the most popular sites according to ``total value locked.''  The total population of DeFi sites is much larger though and our results provide no insight into the privacy and security issues of the long tail of DeFi sites.  Considering the little care and effort that goes into many DeFi sites, we expect the long tail to exhibit more problems than popular sites like Uniswap or Compound.  We therefore believe that our work represents a lower bound of security and privacy issues.  Future work could take a more comprehensive and (perhaps longitudinal) approach to studying the problem.

% There may be selection bias.
Finally, our list of DeFi sites may exhibit \emph{selection bias}, i.e. it may differ from the general population of DeFi sites in crucial aspects. For example, our list of popular DeFi sites may exhibit fewer security and privacy issues than the lesser-known long tail of DeFi sites.

\subsection{Recommendations}
\label{sec:recommendations}

What can DeFi developers do to improve security and privacy for their users, and guarantee the ``De'' in DeFi?  How can users protect themselves?  Below, we issue a set of recommendations for both DeFi developers and users.  For DeFi users, we recommend the following:

\begin{description}
  \item[Block analytics scripts:] 
  To prevent analytics providers from linking Ethereum addresses to real-world identities, we recommend browser extensions like Privacy Badger or browsers like Brave, or Tor Browser.
  \item[Don't connect your wallet unless you have to:] 
  We recommend treating one's Ethereum address like credit card or bank account information, i.e. only revealing it selectively and when necessary.
\end{description}

For developers of DeFi sites, we recommend the following:

\begin{description}
  \item[Use self-hosted analytics:] 
    We recommend using self-hosted analytics scripts instead of third-party services to minimize exposure to third parties.  If a DeFi site does use third party providers, it should ensure that no sensitive information like Ethereum addresses leak to the analytics providers.
  \item[Address privacy as design goal:]
    Our work shows that many DeFi sites don't consider Ethereum addresses private.\footnote{For example, 1inch's referral code contains the inviting user's Ethereum address, which is exposed to the invitee.}  We recommend that DeFi developers treat Ethereum addresses like credit card information.
  \item[Decentralize front ends:] 
    Section~\ref{sec:centralization} highlighted the centralization among DeFi front ends.  To improve reliability in the face of outages and DoS attacks, DeFi site operators should consider making available their front ends over alternative infrastructure such as \textsc{ipfs} or Tor onion services, which also come with desirable security properties such as end-to-end encryption.
  \item[Revise threat models:]  
    Our results show that well-established Web development methods like the use of innocuous support widgets can open the gates to phishing attacks.\footnote{Phishing constitutes a pervasive problem taking on numerous forms, ranging from scammers masquerading as support staff~\cite{wuzz1e2021a} or ``helpful'' social media users~\cite{Young2021a} to fake~\cite{Khatri2019a} or compromised wallet software~\cite{Shevchenko2020a}.}  We recommend that DeFi developers carefully consider what third parties should be allowed to modify their site's \textsc{dom}. Not only does one have to trust that these third parties are \emph{benign}; one also has to trust that they are \emph{competent} and able to secure their infrastructure from compromise---a high bar that may be difficult to meet if substantial amounts of money are at stake.
\end{description}

\iffalse
\subsection{Future Work}
\label{sec:future-work}

Future work could contrast the privacy policies of DeFi sites and focus on questions like: What data is collected and for what purpose?  Is the data shared and if so, with whom?  How long is data retained?

Another topic worth investigating are the HTTP security headers that the underlying Web servers of DeFi sites set.  

\begin{itemize}
    \item Privacy policies.
    \item Security headers.
\end{itemize}
\fi
\section{Related Work}
\label{sec:related}

We conclude this paper by contrasting it with related work, which we divide into online privacy related to third-party tracking and cryptocurrencies, and finally user perception of cryptocurrencies.

\point{Online Privacy and Third-Party Tracking}
Advertisers take advantage of a plethora of ways to track users online~\cite{Acar2014, Englehardt2015, Englehardt2016}.  One of the most pervasive ways to do this is via third party trackers such as Google Analytics, which uses \textsc{http} requests and first party cookies in addition to information about the user's browser and system to compile a profile of a user's online activity~\cite{google-analytics}.

\point{Online Privacy and Security of Cryptocurrencies}
Past work on Bitcoin privacy has mostly focused on the linkability of addresses~\cite{Meiklejohn13a} (along with some exploration of network-layer issues~\cite{Apostolaki2021a}), a PoPETs'18 paper~\cite{Goldfeder18a} by Goldfeder et al. takes a first look at the intersection between Bitcoin and online privacy. An increasing number of online vendors now support payment by Bitcoin but the authors show that online trackers often collect sufficient sensitive information to link a purchase to its subsequent blockchain transaction. Worse, by taking into account auxiliary information, attackers could link together Bitcoin transactions that were anonymized via CoinJoin---a popular mixer at the time. We build on Goldfeder et al.'s first foray into the intersection of cryptocurrency and online privacy, showing that both third and first party scripts can facilitate security and privacy issues in DeFi sites.

A 2020 technical report by Béres et al.~\cite{Beres20a} takes a look at privacy in Ethereum, showing how attackers can profile and deanonymize users. The authors show that one can link several Ethereum addresses to the same owner by taking into account the time of day these addresses are typically used, the gas price, and unrelated addresses that are transacted with. Even mixers like the popular Tornado Cash are no panacea because they are frequently misused. For example, not understanding the nuances of mixers, some users use the same Ethereum address for deposit and withdrawal, effectively deanonymizing themselves. We expand on Béres et al.'s work by pointing out how common Web development methods lead to privacy and security issues in DeFi applications.

Li et al. reveal in their \textsc{ndss}'21 paper~\cite{Li21a} a DoS attack that makes it possible to disable \textsc{rpc} services that DApps rely on---e.g. to get the upper hand in an auction by preventing competing bidders from placing their bids. The attack exploits the fact that many \textsc{rpc} services don’t impose a gas limit on the \texttt{eth\_call} method, making it possible to make the \textsc{rpc} service engage in heavy computation, thus preventing it from serving other clients.  Refer to Werner et al.'s arXiv report~\cite{Werner21a} for a comprehensive overview of DeFi and the state of open research questions around DeFi.

In their~2021 arXiv report, Das et al. study security issues in \textsc{nft}s~\cite{Das2021a} by drawing on data from eight \textsc{nft} market places.  The authors document security issues in market places, fraudulent user behavior, and the ``decay'' in \textsc{nft}s whose content (but not the on-chain reference) disappears over time.

\point{User Perception}
In a FC'16 paper~\cite{Krombholz16a}, Krombholz et al. surveyed 990 Bitcoin users about their understanding of Bitcoin security, privacy, and anonymity. Interestingly, nobody stored wallet backups on an air-gapped computer but several respondents used encrypted backups. 32\% of respondents believe that Bitcoin is per-se anonymous but 80\% think that it is possible to follow their transactions. The study’s respondents were no strangers to loss of Bitcoin: 22\% report that they have lost Bitcoins at least once---due to hardware or software failure, or because they lost access to their private keys. Finally, respondents saw vulnerabilities in hosted wallets as the top threat right after Bitcoin value fluctuation.

User misconceptions go beyond mixers. Drawing on data from twenty interviews with cryptocurrency users and non-users, Voskobojnikov et al. show in their FC’20 paper~\cite{Voskobojnikov20a} that users express confusion about the concept of ``gas prices,'' mistakenly believe that they own the private key for funds stored by the company Coinbase, or don’t understand the idea behind public and private keys altogether. Faulty mental models can lead to critical mistakes like the loss of funds, highlighting the importance of safe defaults that protect users. This work contributes to our understanding of what privacy-preserving safe defaults look like.

Most recently, a \textsc{soups}'20 paper by Mai et al.~\cite{Mai20a} qualitatively studied cryptocurrency users’ (N=29) mental models and how these models are in conflict with security and privacy goals. Corroborating the findings of Voskobojnikov et al., the authors find that the idea of cryptographic keys is a frequent source of misunderstanding, prompting some participants to believe that miners or ``the blockchain'' create private keys. Other participants exhibit misunderstandings about the blockchain, believing that old transactions are eventually deleted, or that transactions are confidential and cannot be seen by third parties. This highlights that some users may have incorrect expectations of privacy.
\section{Conclusions}
\label{sec:conclusions}

This work is the first to study the privacy, security, and decentralization properties of popular DeFi sites.  We conclude that despite the lightweight nature of DeFi front ends, well-understood security and privacy risks are as widespread on DeFi applications as on other parts of the Web---but carry \emph{greater risk} in DeFi given the money that is involved.  The carelessness with which DeFi sites handle users' Ethereum addresses suggests that problems are not only technical but also cultural:  The popular technology ethos of ``move fast and break things'' conflicts with the care that is necessary when processing sensitive financial information.  We also establish that despite claims to the opposite, because of several companies like Amazon and Cloudflare running much of Web's infrastructure, DeFi as a whole is considerably less decentralized than previously believed.
\section*{Availability}

Our data set and source code is available at: \\
\url{https://github.com/brave-experiments/defi-privacy-measurements}

\printbibliography

\appendix

\section{List of URLs}
\label{sec:defi-urls}

\begin{table*}[ht]
\centering
\caption{Our list of 78 DeFi sites.}
\label{tab:defi-urls}
\begin{tabular}{ll}
\toprule
\url{https://activate.codefi.network/staking/airswap/governance} &
\url{https://app.1inch.io} \\
\url{https://app.aave.com/markets} &
\url{https://app.alchemix.fi} \\
\url{https://app.badger.finance} &
\url{https://app.balancer.fi} \\
\url{https://app.bancor.network/eth/data} &
\url{https://app.barnbridge.com} \\
\url{https://app.bifi.finance} &
\url{https://app.bifi.finance/lend} \\
\url{https://app.boringdao.com} &
\url{https://app.compound.finance} \\
\url{https://app.coverprotocol.com} &
\url{https://app.cream.finance} \\
\url{https://app.defisaver.com} &
\url{https://app.dodoex.io} \\
\url{https://app.enzyme.finance/depositor/leaderboard} &
\url{https://app.fei.money} \\
\url{https://app.flexa.network} &
\url{https://app.impermax.finance} \\
\url{https://app.jelly.market} &
\url{https://app.mai.finance} \\
\url{https://app.maple.finance} &
\url{https://app.nexusmutual.io/swap} \\
\url{https://app.pickle.finance} &
\url{https://app.rampdefi.com} \\
\url{https://app.rari.capital} &
\url{https://app.rari.capital} \\
\url{https://app.reflexer.finance} &
\url{https://app.reflexer.finance} \\
\url{https://app.ribbon.finance} &
\url{https://app.rulerprotocol.com} \\
\url{https://app.sushi.com} &
\url{https://app.swapswap.org/#/swap} \\
\url{https://app.tornado.cash} &
\url{https://app.truefi.io} \\
\url{https://app.truefi.io/home} &
\url{https://app.uniswap.org} \\
\url{https://app.vesper.finance} &
\url{https://app.warp.finance} \\
\url{https://app.yield.is} &
\url{https://app.zerion.io} \\
\url{https://beta.curve.fi} &
\url{https://curve.fi} \\
\url{https://dashboard.keep.network/overview} &
\url{https://debank.com} \\
\url{https://defi.instadapp.io} &
\url{https://dmm.exchange/#/about} \\
\url{https://exchange.dfyn.network} &
\url{https://exchange.loopring.io/swap} \\
\url{https://flash.wing.finance} &
\url{https://for.tube/market/index} \\
\url{https://foundation.app} &
\url{https://harvest.finance} \\
\url{https://homora-v2.alphafinance.io} &
\url{https://idle.finance} \\
\url{https://inverse.finance/anchor} &
\url{https://liquity.app} \\
\url{https://moonswap.fi/exchange/swap} &
\url{https://notional.finance/portfolio} \\
\url{https://o3swap.com/swap} &
\url{https://oasis.app/dashboard} \\
\url{https://opensea.io/assets} &
\url{https://pancakeswap.finance} \\
\url{https://pay.sablier.finance} &
\url{https://rarible.com} \\
\url{https://saddle.exchange} &
\url{https://staking.synthetix.io} \\
\url{https://tinlake.centrifuge.io} &
\url{https://trade.dydx.exchange} \\
\url{https://trader.airswap.io/} &
\url{https://v2.opyn.co} \\
\url{https://wasabix.finance/#/app} &
\url{https://www.akropolis.io/app/home} \\
\url{https://www.convexfinance.com} &
\url{https://www.indexcoop.com} \\
\url{https://yearn.finance} &
\url{https://zapper.fi/dashboard} \\
\bottomrule
\end{tabular}
\end{table*}

\end{document}